\documentclass[12pt]{article}
\newcommand{\beq}{\begin{equation}}
\newcommand{\eeq}{\end{equation}}
\newcommand{\bra}{\begin{array}}
\newcommand{\era}{\end{array}}

\newcommand{\al}{\alpha}

\newcommand{\de}{\delta}

\newcommand{\ep}{\epsilon}
\let\include\input
\input epsf.sty
\usepackage{latexsym}
\author{Jamila Douari\footnote{jdouari@excite.com} and Arafa H. Ali\footnote{arafa16@yahoo.com}\\ \\
\small\it Center for Advanced Mathematical Sciences\\\small\it
American University of Beirut\\ \small\it P.O.Box 11-0236, College
Hall \\\small\it Beirut, Lebanon\rm }
\title{Funnel's Fluctuations in Dyonic Case: Intersecting D1-D3 Branes}
\begin{document}
\maketitle \vspace*{0.5cm} PACS: 11.25.-w, 11.25.Uv \vskip1cm
Keywords: Branes, Dyons, Fluctuations, Boundary Conditions.
\vspace*{1.5cm}
\section*{Abstract}
\hspace{.3in}The fluctuations of funnel solutions of intersecting D1
and D3 branes are quite explicitly discussed by treating different
modes and different directions of the fluctuation at the presence of
world volume electric field. The boundary conditions are found to be
Neumann boundary conditions.
\section{Introduction}
\hspace{.3in}D-branes described by Non-abelian Born-Infeld (BI) action \cite{BI} have
many fascinating features. Among these there is the possibility for
D-branes to morph into other D-branes of different dimensions by
exciting some of the scalar fields \cite{InterBran1,InterBran2}.
It's known in the literature that there are many different but
physically equivalent descriptions of how a D1-brane may end on a
D3-brane. From the point of view of the D3 brane the configuration
is described by a monopole on its world volume. From the point of
view of the D1-brane the configuration is described by the D1-brane
opening up into a D3-brane where the extra three dimensions form a
fuzzy two-sphere whose radius diverges at the origin of the
D3-brane. These different view points are the stringy realization of
the Nahm transformation \cite{fun,dual}. Also the dynamics of the
both bion spike \cite{InterBran1,fluct} and the fuzzy funnel
\cite{dual,fuzfun,funnelSoluD5} were studied by considering
linearized fluctuations around the static solutions.\\

The present work is devoted to study the fluctuations of funnel
solutions in the presence of a world-volume electric field. By
discussing the solutions and the potentials for this particular case
we end by the system D1$\bot$D3 branes gets a special property
because of the presence of electric field; the system is divided to
two regions corresponding to small and large electric field.
Consequently, the system has Neumann boundary conditions and the end
of open string can move freely on the brane which is agree with its
dual discussed in \cite{NBCBI} considering Born-Infeld action
dealing with the fluctuation of the bion skipe in D3$\bot$D1-case.\\

The paper is organized as follows: In section 2, we start by a brief
review  on D1$\bot$D3 branes in dyonic case by using the
non-Born-Infeld action. Then, we discuss the fluctuations of the
fuzzy funnel in section 3 for zero and high modes. We give the
solutions of the linearized equations of motion of the fluctuations
for both cases the overall transverse and the relative one. We also
discuss the solutions and the potential depending on the presence of
electric field which is leading to Neumann boundary conditions as
special property of the system. Then the waves on the brane cause
the fuzzy funnel to freely oscillate.

\section{D1$\bot$D3 Branes with Electric Field Swished On}
\hspace{.3in}In this section, we review in brief the funnel
solutions for D1$\bot$D3 branes from D3 and D1 branes points of
view. First, using abelian BI action for the world-volume gauge
field and one excited transverse scalar in dyonic case, we give the
funnel solution. It was showed in \cite{cm} that the BI action, when
taken as the fundamental action, can be used to build a
configuration with a semi-infinite fundamental string ending on a
D3-brane \cite{Gib}. The dyonic system is given by using D-string
world-volume theory and the fundamental strings introduced by
adding a $U(1)$ electric field. Thus the system is described by the
following action \beq\bra{lll} S=\int dt L &=-T_3 \int
d^4\sigma\sqrt{-det(\eta_{ab}+\lambda^2
\partial_a \phi^i \partial_b \phi^i +\lambda F_{ab})}\\\\
&= -T_3 \int d^4\sigma\Big[ 1 +\lambda^2 \Big( \mid \nabla\phi \mid^2 +B^2 +E^2 \Big)\\\\
&+\lambda^4 \Big( (B.\nabla\phi )^2 +(E.B)^2 +\mid E
\wedge\nabla\phi\mid^2 \Big) \Big]^{\frac{1}{2}} \era\eeq in which
$F_{ab}$ is the field strength and the electric field is denoted as
$F_{09}=EI_{ab}$, ($I_{ab}$ is $N\times N$ matrix). $\sigma^a$
($a=0,...,3$) denote the world volume coordinates while $\phi^i$
($i=4,...,9$) are the scalars describing transverse fluctuations of
the brane and $\lambda=2\pi \ell_s^2$ with $\ell_s$ is the string
length. In our case we excite just one scalar so $\phi^i=\phi^9
\equiv\phi$. Following the same process used in the reference
\cite{cm} by considering static gauge, we look for the lowest energy
of the system. Accordingly to (1) the energy of dyonic system is
given as \beq\bra{ll} \Xi&= T_3 \int d^3\sigma\Big[ \lambda^2 \mid
\nabla\phi +\stackrel{\rightarrow}{B}
+\stackrel{\rightarrow}{E}\mid^2 +(1-\lambda^2
\nabla\phi.\stackrel{\rightarrow}{B})^2-2\lambda^2
\stackrel{\rightarrow}{E}.(\stackrel{\rightarrow}{B}
+\nabla\phi)\\\\
&+ \lambda^4 \Big(
(\stackrel{\rightarrow}{E}.\stackrel{\rightarrow}{B})^2 +\mid
\stackrel{\rightarrow}{E} \Lambda\nabla\phi\mid^2\Big)
\Big]^{1/2}.\era\eeq Then if we require $\nabla\phi
+\stackrel{\rightarrow}{B} +\stackrel{\rightarrow}{E}=0$, $\Xi$
reduces to $\Xi_0\geq0$ and we find \beq\bra{ll} \Xi_0 &=T_3 \int
d^3\sigma\Big[(1-\lambda^2
(\nabla\phi).\stackrel{\rightarrow}{B})^2+2\lambda^2
\stackrel{\rightarrow}{E}.\stackrel{\rightarrow}{E} )\\\\&+
\lambda^4 ( (\stackrel{\rightarrow}{E}.\stackrel{\rightarrow}{B})^2
+{\mid\stackrel{\rightarrow}{E}}\Lambda \nabla\phi\mid^2)
\Big]^{1/2}\era\eeq as minimum energy. By using the Bianchi identity
$ \nabla.B= 0$ and the fact that the gauge field is static, the
funnel solution is then \beq\phi =\frac{N_m +N_e}{2r},\eeq
with $N_m$ is magnetic charge and $N_e$ electric charge.\\

Now we consider the dual description of the $D1\bot D3$ from D1
branes point of view. To get D3-branes from D-strings, we use the
non-abelian BI action \beq S=-T_1\int d^2\sigma STr \Big[
-det(\eta_{ab}+\lambda^2 \partial_a \phi^i Q_{ij}^{-1}\partial_b
\phi^j )det Q^{ij}\Big]^{1\over2}\eeq where $ Q_{ij}=\de_{ij}
+i\lambda \lbrack \phi_i , \phi_j \rbrack$. Expanding this action to
leading order in $\lambda$ yields the usual non-abelian scalar
action
$$S\cong -T_1\int d^2\sigma  \Big[ N+ \lambda^2 Tr (\partial_a
\phi^i + \frac{1}{2}\lbrack \phi_i , \phi_j \rbrack \lbrack \phi_j ,
\phi_i \rbrack) +...\Big]^{1\over2}.$$ The solutions of the equation
of motion of the scalar fields $\phi_i$, $i=1,2,3$ represent the
D-string expanding into a D3-brane analogous to the bion solution of
the D3-brane theory \cite{InterBran1,InterBran2}. The solutions are
$$\bra{lc}\phi_i =\pm\frac{\al_i}{2\sigma},&\lbrack \al_i , \al_j
\rbrack=2i\ep^{ijk}\al_k ,\era$$ with the corresponding geometry is
a long funnel where the cross-section at fixed $\sigma$ has the
topology of a fuzzy two-sphere.\\

The dyonic case is taken by considering ($N, N_f$)-strings. We have
$N$ D-strings and $N_f$ fundamental strings \cite{dual}. The theory
is described by the action \beq S=-T_1\int d^2\sigma STr \Big[
-det(\eta_{ab}+\lambda^2 \partial_a \phi^i Q_{ij}^{-1}\partial_b
\phi^j +\lambda EI_{ab})det Q^{ij}\Big]^{1\over2}\eeq in which we
replaced the field strength $F_{ab}$ by $EI_{ab}$ ($I_{ab}$ is
$N\times N$-matrix) meaning that
the fundamental string is introduced by adding a $U(1)$ electric field $E$.\\

The action can be rewritten as \beq S=-T_1\int d^2\sigma STr \Big[
-det\pmatrix{\eta_{ab}+\lambda EI_{ab}& \lambda \partial_a \phi^j
\cr -\lambda \partial_b \phi^i & Q^{ij}\cr}\Big]^{1\over2}.\eeq Then
the bound states of D-strings and fundamental strings are made
simply by introducing a background $U(1)$ electric field on
D-strings, corresponding to fundamental strings dissolved on the
world-sheet. By computing the determinant, the action becomes \beq
S=-T_1\int d^2\sigma STr \Big[ (1-\lambda^2 E^2 + \al_i \al_i
\hat{R}'^2)(1+4\lambda^2 \al_j \al_j \hat{R}^4 )\Big]^{1\over2}.\eeq
where the following ansatz were inserted \beq\phi_i =\hat{R}\al_i
.\eeq Hence, we get the funnel solution for dyonic string by solving
the equation of variation of $\hat{R}$, as follows \beq \phi_i
=\frac{\al_i}{2\sigma\sqrt{1-\lambda^2 E^2}} .\eeq

\section{Fluctuations of Dyonic Funnel Solutions}
\hspace{.3in}In this section, we treat the dynamics of the funnel
solutions. We solve the linearized equations of motion for small and
time-dependent fluctuations of the transverse scalar around the
exact background in dyonic case.\\

We deal with the fluctuations of the funnel (10) discussed in the
previous section. By plugging into the full ($N-N_f$) string action
(6,7) the "overall transverse" $\delta \phi^m (\sigma,t)=f^m
(\sigma,t)I_N$, $m=4,...,8$ which is the simplest type of
fluctuation with $I_N$ the identity matrix, together with the funnel
solution, we get \beq\bra{llll} S&=-T_1\int d^2\sigma STr \Big[
(1+\lambda E)(1+\frac{\lambda^2 \al^i \al^i}{4\sigma^4 }) \Big(
(1+\frac{\lambda^2 \al^i \al^i}{4\sigma^4})(1+(\lambda E
-1)\lambda^2 (\partial_t \delta\phi^m )^2) +\lambda^2
(\partial_\sigma \delta\phi^m )^2\Big)
\Big]^{1\over 2}\\\\
&\approx-NT_1\int d^2\sigma H \Big[ (1+\lambda E)-(1-\lambda^2
E^2)\frac{\lambda^2}{2} (\dot{f}^m)^2 +\frac{(1+\lambda
E)\lambda^2}{2H} (\partial_\sigma f^m)^2 +...\Big] \era \eeq where
$$H=1+\frac{\lambda^2 C}{4\sigma^4}$$ and $C=Tr \al^i \al^i$. For
the irreducible $N\times N$ representation we have $C=N^2 -1$. In
the last line we have only kept the terms quadratic in the
fluctuations as this is sufficient to determine the linearized
equations of motion \beq\Big( (1-\lambda E)(1+\lambda^2\frac{N^2
-1}{4\sigma^4})\partial^{2}_{t}-\partial^{2}_{\sigma}\Big) f^m
=0.\eeq

In the overall case, all the points of the fuzzy funnel move or
fluctuate in the same direction of the dyonic string by an equal
distance $\delta x^m$. First, the funnel solution is $\phi^i
=\frac{1}{2\sqrt{1-\lambda^2 E^2}}\frac{\al^i}{\sigma}$ and the
fluctuation $f^m$ waves in the direction of $x^m$; \beq f^m
(\sigma,t)=\Phi(\sigma)e^{-iwt}\delta x^m.\eeq With this ansatz the
equation of motion is \beq\Big( (1-\lambda E)H w^2
+\partial^{2}_{\sigma}\Big) \Phi(\sigma)=0.\eeq Then, the problem is
reduced to finding the solution of a single scalar equation.\\

Thus, we remark that the equation (14) is an analog one-dimensional
Schr\"odinger equation and it can be rewritten as \beq\Big(
-\partial^{2}_{\sigma}+V(\sigma)\Big) \Phi(\sigma)=w^2(1-\lambda
E)\Phi(\sigma),\eeq with
$$V(\sigma)=w^2(\lambda E-1)\lambda^2\frac{N^2 -1}{4\sigma^4}.$$
We notice that, if the electric field dominates $E\gg 1$, the
potential goes to $w^2\lambda^3 E\frac{N^2}{4\sigma^4}$ for large
$N$ and if $E\ll 1$ we find $V= -w^2 \lambda^2
\frac{N^2}{4\sigma^4}$. This can be seen as two separated systems
depending on electric field so we have Neumann boundary condition
separating the system into two regions $E\gg 1$ and $E\ll 1$.\\

Now, let's find the solution of a single scalar equation (14).
First, the equation (14) can be rewritten as follows \beq\Big(
\frac{1}{w^2(1-\lambda E)}\partial^{2}_{\sigma}+1+\frac{\lambda^2
N^2 }{4\sigma^4}\Big) \Phi(\sigma)=0,\eeq for large $N$. If we
suggest $\tilde{\sigma}=w\sqrt{1-\lambda E}\sigma$ the latter
equation becomes \beq\Big(
\partial^{2}_{\tilde{\sigma}}+1+\frac{\kappa^2}{\tilde{\sigma}^4}\Big)
\Phi(\tilde{\sigma})=0,\eeq with the potential is \beq
V(\tilde{\sigma})=\frac{\kappa^2}{\tilde{\sigma}^4}, \eeq and
$\kappa=\frac{\lambda N w^2}{2}(1-\lambda E)$. This equation is a
Schr\"odinger equation for an attractive singular potential
$\propto\tilde{\sigma}^{-4}$ and depends on the single coupling
parameter $\kappa$ with constant positive Schr\"odinger energy. The
solution is then known by making the following coordinate change
\beq
\chi(\tilde{\sigma})=\int\limits^{\tilde{\sigma}}_{\sqrt{\kappa}}
dy\sqrt{1+\frac{\kappa^2}{y^4}}, \eeq and \beq
\Phi=(1+\frac{\kappa^2}{\tilde{\sigma}^4})^{-\frac{1}{4}}\tilde{\Phi}.
\eeq Thus, the equation (17) becomes \beq\Big(
-\partial^{2}_{\chi}+V(\chi)\Big) \tilde{\Phi}=0,\eeq with \beq
V(\chi)=\frac{5\kappa^2}{(\tilde{\sigma}^2
+\frac{\kappa^2}{\tilde{\sigma}^2})^3}.\eeq Then, the fluctuation is
found to be \beq
\Phi=(1+\frac{\kappa^2}{\tilde{\sigma}^4})^{-\frac{1}{4}}e^{\pm
i\chi(\tilde{\sigma})}. \eeq This fluctuation has
the following limits; at large $\sigma$, $\Phi\sim e^{\pm
i\chi(\tilde{\sigma})}$ and if $\sigma$ is small
$\Phi=\frac{\sqrt{\kappa}}{\tilde{\sigma}}e^{\pm
i\chi(\tilde{\sigma})}$. These are the asymptotic
wave function in the regions $\chi\rightarrow \pm\infty$, while
around $\chi\sim 0$; i.e. $\tilde{\sigma}\sim\sqrt{\kappa}$,
$f^m\sim 2^{-\frac{1}{4}}e^{-iwt}\delta x^m$ (Fig.1).\\

The potential (22) in large and small limits of electric field becomes (Fig.2);\\
\begin{itemize}
\item $E\gg 1$, $V(\chi)\sim \frac{-5\lambda N^2}{ E\sigma^6}$
\item $E\ll 1$, $V(\chi)\sim \frac{5\lambda^2 N^2 w^2}{4(w^2 \sigma^2
+\frac{\lambda^2 N^2 w^2}{4\sigma^2})}$
\end{itemize}
At the presence of electric field we remark that around $\sigma\sim 0$ there is a symmetric potential which goes to zero very fast and more fast as electric field is large $\sim\frac{-1}{E\sigma^2}$. As discussed above, again we get the separated systems in different
regions depending on the values of electric field. Also if we have a
look at the fluctuation (23) we find that $f^m$ in
the case of $E\gg 1$ is different from the one in $E\ll 1$ case and as shown in the fig.1 the presence of electric field causes a discontinuity of the fluctuation wave which means free boundary condition. Contrarily, at the absence of electric field the fluctuation wave is continue.
Then, this is seen as Neumann boundary condition from
non-Born-Infeld dynamics separating the system into two regions
$E\gg 1$ and $E\ll 1$ which is agree with its dual discussed in
\cite{NBCBI}. \\
\begin{figure}
\begin{center}
\mbox{\epsfysize=6cm
  \epsfxsize=6cm
  \epsffile{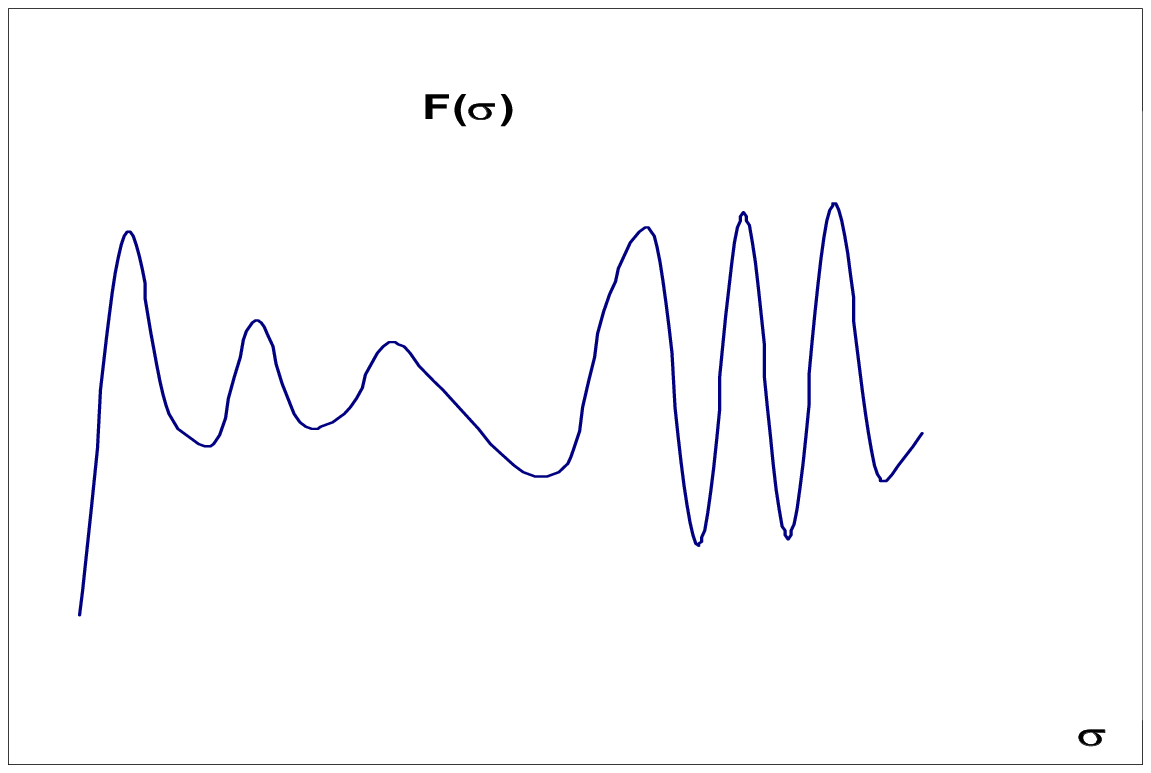}}
\mbox{\epsfysize=6cm
   \epsfxsize=6cm
  \epsffile{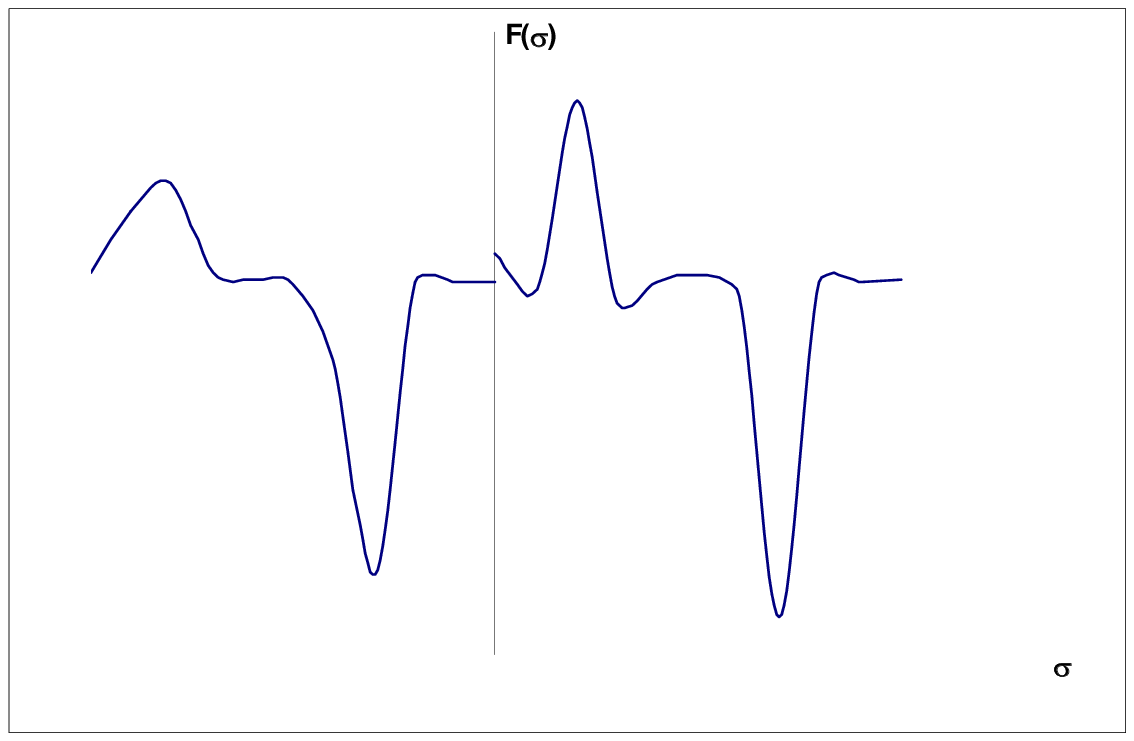}}
\end{center}
\bf{Figure 1 : }\rm Left hand curve represents the overall fluctuation wave in zero mode and low electric field. Right hand curve shows the scattering of the overall fluctuation wave in zero mode and high electric field. This latter caused a discontinuitity of the wave which means Neumann boundary condition.
\end{figure}
\begin{figure}
\begin{center}
\mbox{\epsfysize=5cm
   \epsfxsize=5cm
  \epsffile{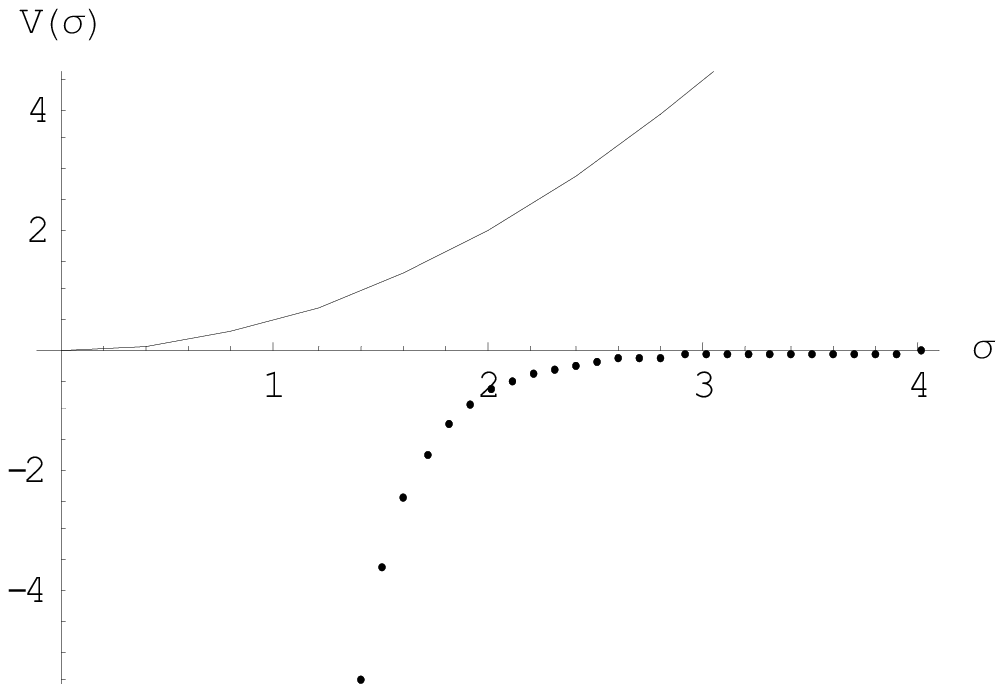}}
\end{center}
\bf{Figure 2 : }\rm The up line shows the potential in zero mode of the overall funnel's fluctuations at the absence of electric field E and the dots represent the potential in the same mode at the presence of E. The presence of E is changing the potential totally to the opposite.
\vskip 1cm
\end{figure}
The fluctuations discussed above could be called the zero mode
$\ell=0$ and for high modes $\ell\geq0$, the fluctuations are
$\delta \phi^m (\sigma,t)=\sum\limits^{N-1}_{\ell=0}\psi^{m}_{i_1
... i_\ell}\al^{i_1} ... \al^{i_\ell} $ with $\psi^{m}_{i_1 ...
i_\ell}$ are completely symmetric and traceless in the lower
indices.\\
The action describing this system is \beq\bra{lll}
S&\approx-NT_1\int d^2\sigma  \Big[ (1+\lambda E)H-(1-\lambda^2
E^2)H\frac{\lambda^2}{2} (\partial_{t}\delta\phi^m)^2) \\\\
&+\frac{(1+\lambda E)\lambda^2}{2H} (\partial_\sigma \delta\phi^m)^2
-(1-\lambda^2 E^2)\frac{\lambda^2}{2}\lbrack \phi^i ,\delta\phi^m
\rbrack^2 \\\\
&-\frac{\lambda^4}{12}\lbrack \partial_{\sigma}\phi^i
,\partial_{t}\delta\phi^m \rbrack^2+...\Big] \era\eeq Now the
linearized equations of motion are \beq \Big[(1+\lambda
E)H\partial_{t}^2 -\partial_{\sigma}^2\Big]\delta\phi^m
+(1-\lambda^2 E^2)\lbrack \phi^i ,\lbrack \phi^i ,\delta\phi^m
\rbrack\rbrack -\frac{\lambda^2}{6}\lbrack \partial_{\sigma}\phi^i
,\lbrack \partial_{\sigma}\phi^i ,\partial^2 _{t}\delta\phi^m
\rbrack\rbrack=0.\eeq Since the background solution is $\phi^i
\propto \al^i$ and we have $\lbrack \al^i , \al^j \rbrack
=2i\epsilon_{ijk}\al^k $, we get \beq\bra{ll} \lbrack \al^i ,
\lbrack\al^i, \delta\phi^m \rbrack
&=\sum\limits_{\ell<N}\psi^{m}_{i_1 ... i_\ell}\lbrack \al^i ,
\lbrack\al^i ,\al^{i_1} ... \al^{i_\ell} \rbrack\\\\
&=\sum\limits_{\ell<N}4\ell(\ell+1)\psi^{m}_{i_1 ...
i_\ell}\al^{i_1} ... \al^{i_\ell} \era\eeq To obtain a specific
spherical harmonic on 2-sphere, we have \beq\lbrack \phi^i ,\lbrack
\phi^i ,\delta\phi_{\ell}^m
\rbrack\rbrack=\frac{\ell(\ell+1)}{\sigma^2}\delta\phi_{\ell}^m
,\phantom{~~~~~~}\lbrack \partial_{\sigma}\phi^i ,\lbrack
\partial_{\sigma}\phi^i ,\partial_{t}^2 \delta\phi^m
\rbrack\rbrack=\frac{\ell(\ell+1)}{\sigma^4}\partial_{t}^2\delta\phi
_{\ell}^m .\eeq Then for each mode the equations of motion are \beq
\Big[ \Big( (1+\lambda E)(1+\lambda^2\frac{N^2 -1}{4\sigma^4})
-\frac{\lambda^2\ell(\ell+1)}{6\sigma^4}\Big) \partial_{t}^2
-\partial_{\sigma}^2 +(1-\lambda^2 E^2)\frac{\ell(\ell+1)}{\sigma^2}
\Big]\delta\phi_{\ell}^m =0.\eeq The solution of the equation of
motion can be found by taking the following proposal. Let's consider
$\phi_{\ell}^m =f^m_\ell (\sigma)e^{-iwt}\delta x^m$ in direction
$m$ with $f^m_\ell (\sigma)$ is some function of $\sigma$ for each
mode
$\ell$.\\

The last equation can be rewritten as \beq \Big[-\partial_{\sigma}^2
+V(\sigma) \Big] f_{\ell}^m (\sigma)=w^2 (1+\lambda E) f_{\ell}^m
(\sigma),\eeq with
$$V(\sigma)=-w^2 \Big( (1+\lambda E)\frac{\lambda^2 N^2}{4\sigma^4}
-\frac{\lambda^2\ell(\ell+1)}{6\sigma^4}\Big) +(1-\lambda^2
E^2)\frac{\ell(\ell+1)}{\sigma^2}.$$

Let's write the equation (29) in the following form \beq \Big[ w^2 \Big( (1+\lambda E)H
-\frac{\lambda^2\ell(\ell+1)}{6\sigma^4}\Big) -(1-\lambda^2
E^2)\frac{\ell(\ell+1)}{\sigma^2}+\partial_{\sigma}^2 \Big]
f_{\ell}^m  (\sigma)= 0.\eeq and again as \beq \Big[ 1+
\frac{1}{\sigma^4}\Big(\lambda^2\frac{N^2
-1}{4}-\frac{\lambda^2\ell(\ell+1)}{6(1+\lambda E)}\Big) -(1-\lambda
E)\frac{\ell(\ell+1)}{w^2\sigma^2}+\frac{1}{w^2 (1+\lambda
E)}\partial_{\sigma}^2  \Big] f_{\ell}^m  (\sigma)= 0.\eeq

We define new coordinate $\tilde{\sigma}=w\sqrt{1+\lambda E}\sigma$
and the latter equation becomes \beq
\Big[\partial_{\tilde{\sigma}}^2 + 1+
\frac{\kappa^2}{\tilde{\sigma}^4}+\frac{\eta}{\tilde{\sigma}^2}
\Big] f_{\ell}^m (\sigma)= 0,\eeq where
$$\bra{lr}
\kappa^2=w^2 (1+\lambda E) \Big(\lambda^2\frac{N^2
-1}{4}-\frac{\lambda^2\ell(\ell+1)}{6(1+\lambda
E)}\Big)^{\frac{1}{2}} ,& \eta=-(1-\lambda^2 E^2)\ell(\ell+1) \era
$$
such that $$N>\sqrt{\frac{2\ell(\ell+1)}{3(1+\lambda E)}+1}.$$ For
simplicity we choose small $\sigma$, then the equation (32) is
reduced to \beq \Big[\partial_{\tilde{\sigma}}^2 + 1+
\frac{\kappa^2}{\tilde{\sigma}^4} \Big] f_{\ell}^m (\sigma)= 0,\eeq
as we did in zero mode, we get the solution by using the steps
(19-22) with new $\kappa$. Since we considered small $\sigma$ we get
$$
V(\chi)=\frac{5\tilde{\sigma}^6}{\kappa^4}.
$$
Then \beq f^m_\ell=\frac{\tilde{\sigma}}{\sqrt{\kappa}}e^{\pm
i\chi(\tilde{\sigma})}.\eeq This fluctuation has two different
values at large $E$ and small $E$ (Fig.3) and a closer look at the
potential at large and fixed $N$ in large and small limits of electric field
leads to
\begin{itemize}
\item $E\gg 1$, $V(\chi)\sim  \frac{20w^2 E\sigma^6}{\lambda N^2}$,
\item $E\ll 1$, $V(\chi)\sim \frac{5w^2 \sigma^6}{\lambda^2 \Big(\frac{N^2
}{4}-\frac{\ell(\ell+1)}{6}\Big)}$
\end{itemize}
The potential in the first case is going fast to infinity than the one in the second case because of the electric field if $\sigma\ll 1$ (Fig.4).\\

For large $\sigma$ the equation of motion (30) of the fluctuation
becomes \beq \Big[-\partial_{\sigma}^2 +\tilde{V}(\sigma) \Big]
f_{\ell}^m (\sigma)= w^2 (1+\lambda E) f_{\ell}^m (\sigma),\eeq with
$\tilde{V}(\sigma)=  \frac{(1-\lambda^2 E^2
)\ell(\ell+1)}{\sigma^2}$ and $f_{\ell}^m$ is now a Sturm-Liouville
eigenvalue problem (Fig.3). We found that the fluctuation has discontinuity at the presence of electric field meaning free boundary condition. Also we remark that the potential has different values in the different regions of electric field $E\gg 1$ and $E\ll 1$ and this time for large $\sigma$. In this side, the potential drops with opposite sign from one case to other and as shown in (fig.4). The presence of E is changing the potential totally to the opposite in both cases zero and high modes.
\begin{figure}
\begin{center}
\mbox{\epsfysize=6cm
   \epsfxsize=6cm
  \epsffile{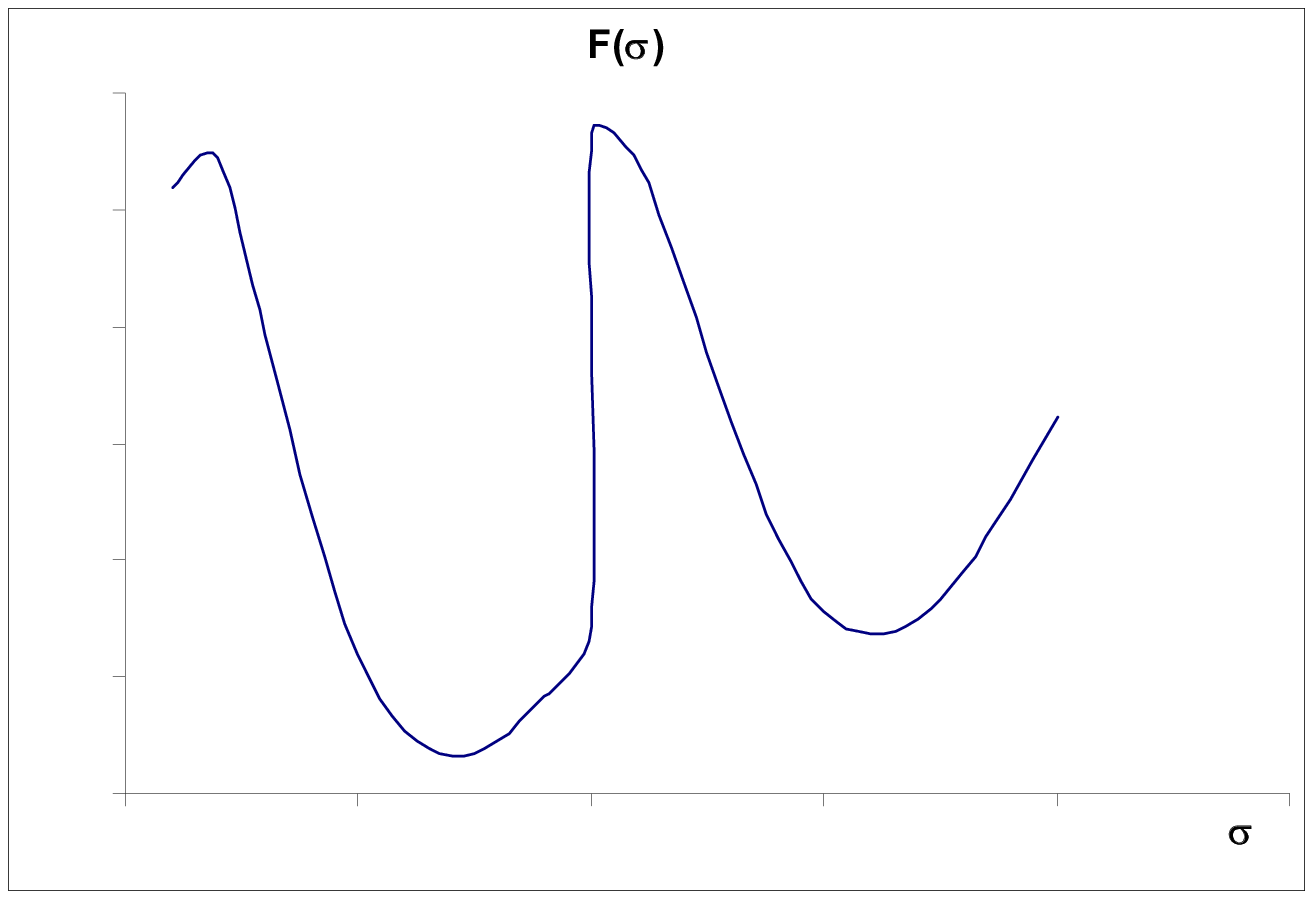}}
\mbox{\epsfysize=6cm
   \epsfxsize=6cm
  \epsffile{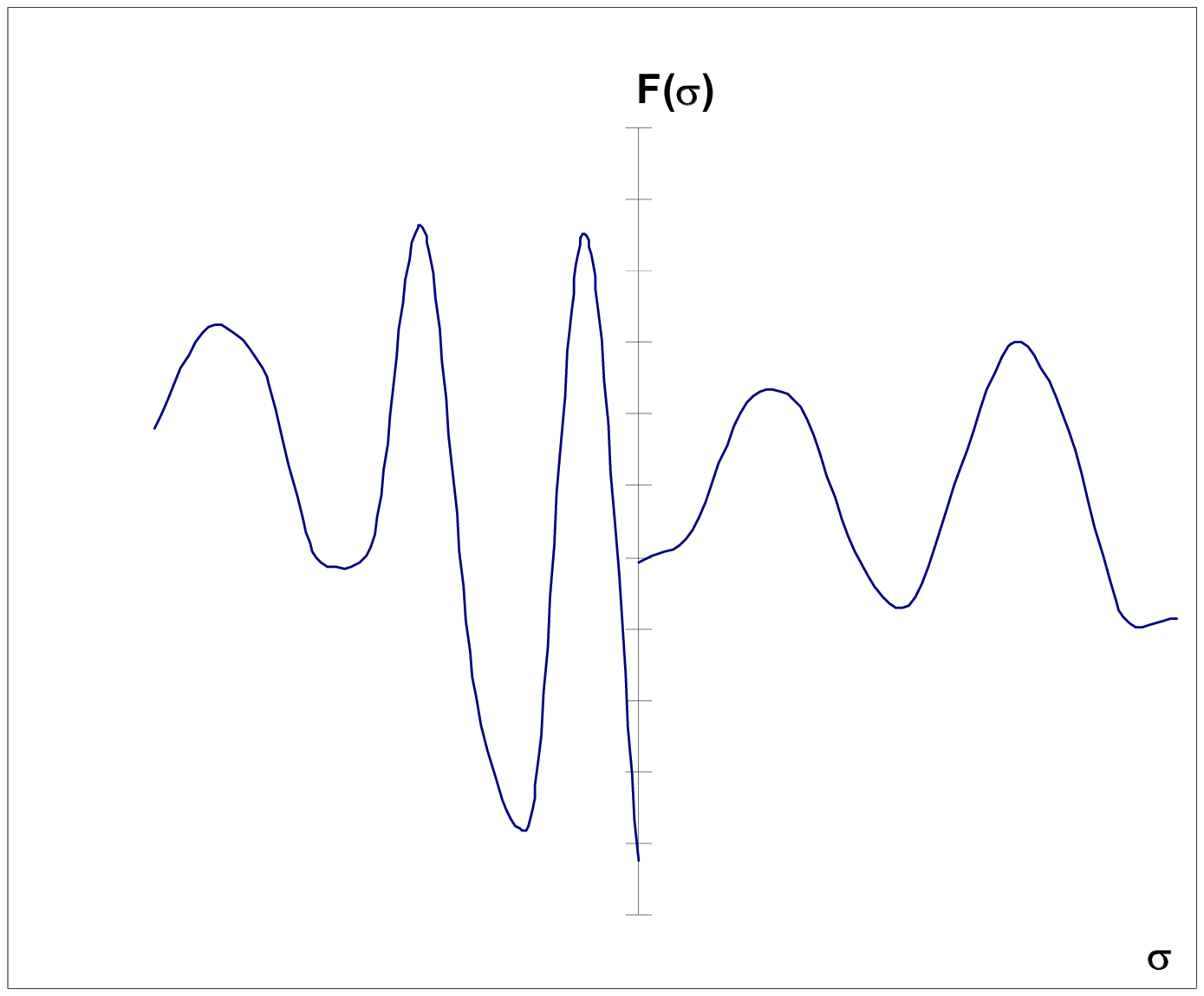}}
\end{center}
\bf{Figure 3 : }\rm The left figure shows the continuity of the fluctuation wave in high mode of the overall fluctuation at the absence of electric field E. The right figure shows the discontinuity of the wave at the presence of E in high mode meaning free boundary condition.
\vskip 1cm
\end{figure}

\begin{figure}
\begin{center}
\mbox{\epsfysize=6cm
   \epsfxsize=6cm
  \epsffile{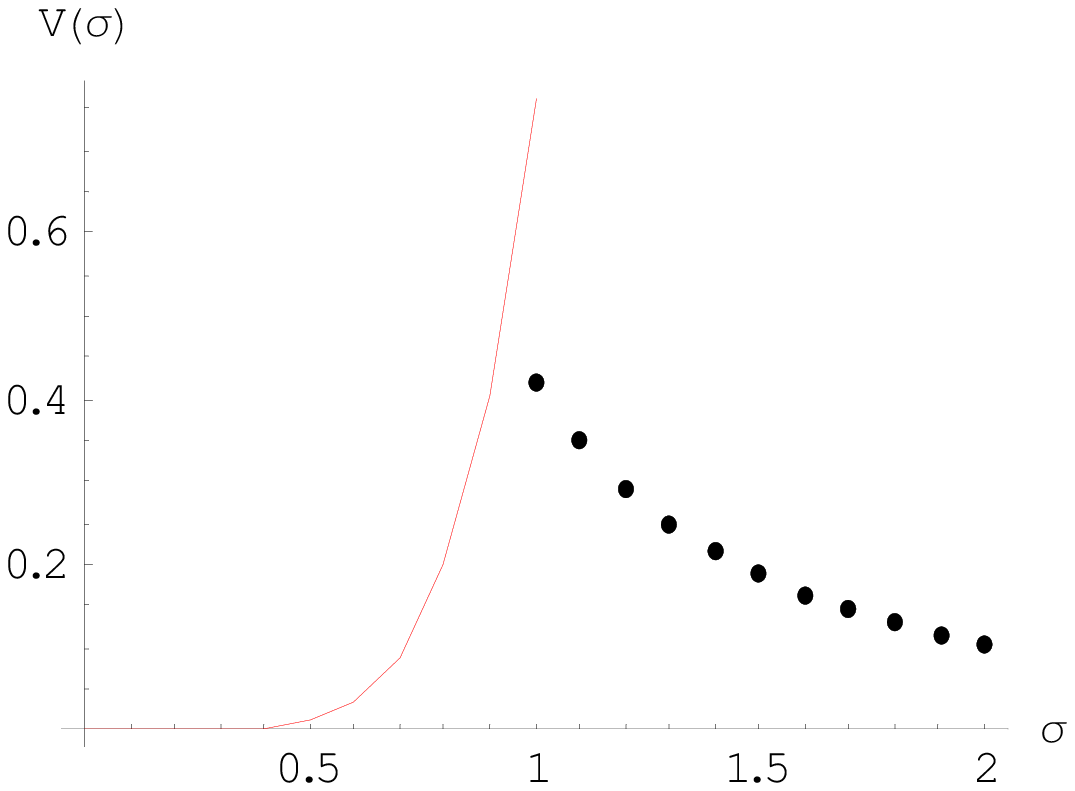}}
\mbox{\epsfysize=6cm
   \epsfxsize=6cm
  \epsffile{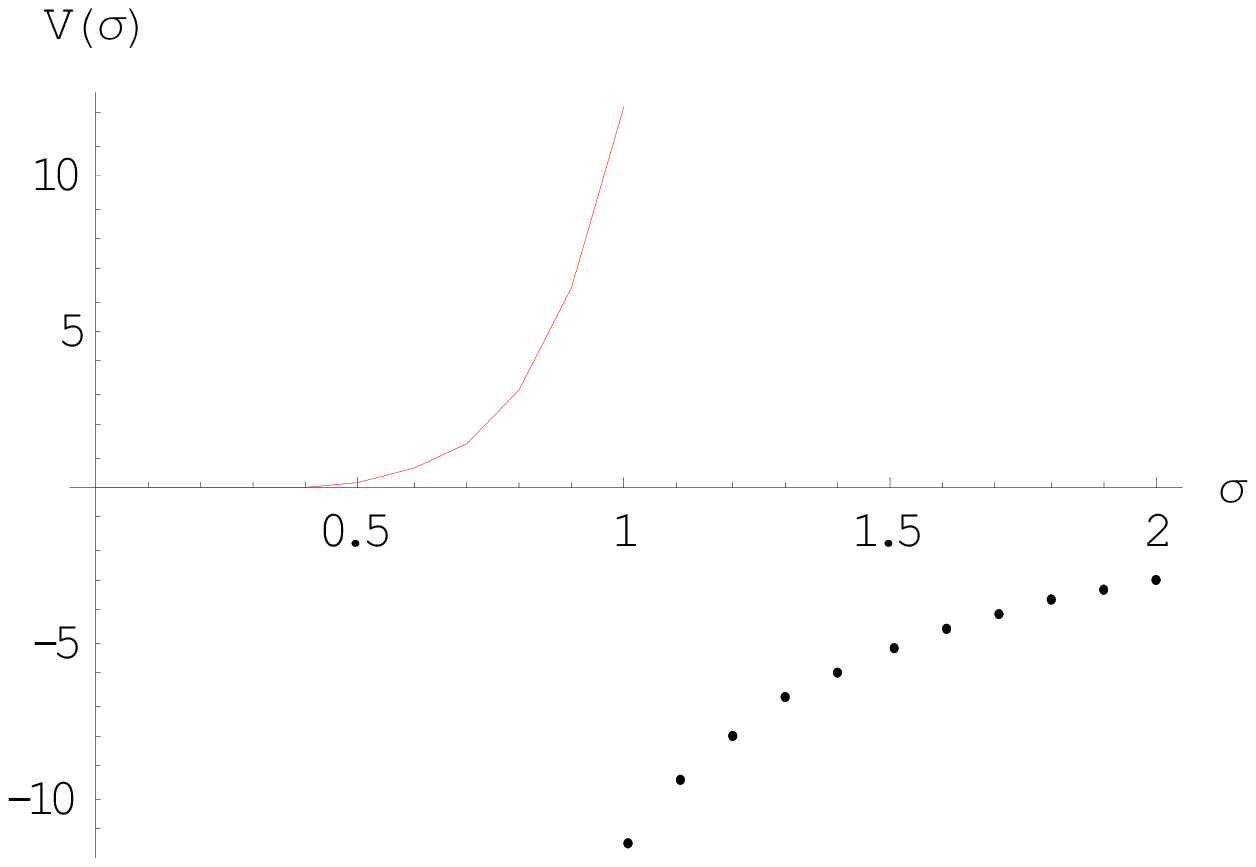}}
\end{center}
\vskip 1cm
\bf{Figure 4 : }\rm The line represents the potential for small $\sigma$ and dots for large $\sigma$ in both figures. In high mode of overall fluctuations at the absence of electric field E, the left figure shows high potential at some stage of $\sigma$ where the two curves meet. The right figure shows a critical case. The curves represent the potentials at the presence of E for small and large $\sigma$.  As a remark, there is no intersecting point for theses two potentials! At some stage of $\sigma$ there is a singularity.
\vskip 1cm
\end{figure}

Consequently, by discussing explicitly the fluctuations and the
potential of intersecting D1-D3 branes in D1-brane world volume
theory we found that the system has Neumann boundary conditions and
the end of the string can move freely on the brane for both zero and
high modes of the overall transverse fluctuations case.
\subsection{Relative Transverse Fluctuations}
\hspace{.3in}Now if we consider the "relative transverse" $\delta
\phi^i (\sigma,t)=f^i (\sigma,t)I_N$, $i=1,2,3$ the action is \beq
S=-T_1\int d^2\sigma STr \Big[ -det\pmatrix{\eta_{ab}+\lambda
EI_{ab}& \lambda \partial_a (\phi^j +\delta\phi^j )\cr -\lambda
\partial_b (\phi^i +\delta\phi^i )& Q^{ij}_*
\cr}\Big]^{1\over2},\eeq with $Q^{ij}_* =Q^{ij}+i\lambda (\lbrack
\phi_i,\delta\phi_j \rbrack+\lbrack \delta\phi_i,\phi_j
\rbrack+\lbrack \delta\phi_i,\delta\phi_j \rbrack)$. As before we
keep only the terms quadratic in the fluctuations and the action
becomes \beq S\approx-NT_1\int d^2\sigma  \Big[ (1-\lambda^2
E^2)H-(1-\lambda E)\frac{\lambda^2}{2}
(\dot{f}^i)^2+\frac{(1+\lambda E)\lambda^2}{2H} (\partial_\sigma
f^i)^2 +...\Big] .\eeq Then the equations of motion of the
fluctuations are \beq\Big ( -\partial^{2}_{\sigma}-w^2
\frac{1-\lambda E}{1+\lambda E}\lambda^2\frac{N^2 -1}{4\sigma^4}\Big
)f^i = w^2 \frac{1-\lambda E}{1+\lambda E}f^i.\eeq

If we write $f^i =\Phi^i (\sigma)e^{-iwt}\de x^i$ in the direction
of $x^i$, the potential will be
$$V(\sigma)=-\frac{1-\lambda E}{1+\lambda E}\lambda^2\frac{N^2
-1}{4\sigma^4}w^{2}.$$ Let's discuss the cases of electric field;
\begin{itemize} \item $E\ll 1$, $V(\sigma)\sim -\lambda^2\frac{N^2
-1}{4\sigma^4}w^{2}$ \item $E\gg 1$, $V(\sigma)\sim
\lambda^2\frac{N^2 -1}{4\sigma^4}w^{2}$
\end{itemize}
Also in the relative case, this is Neumann boundary condition
(Fig.5) which can be also shown by finding the solution of (38) for
which we follow the same way as above by making a coordinate change
suggested by WKB. This case is seen as a zero mode of what is
following so we will treat this in general case by using this
coordinate change for
high modes.
\begin{figure}
\begin{center}
\mbox{\epsfysize=8cm
   \epsfxsize=8cm
  \epsffile{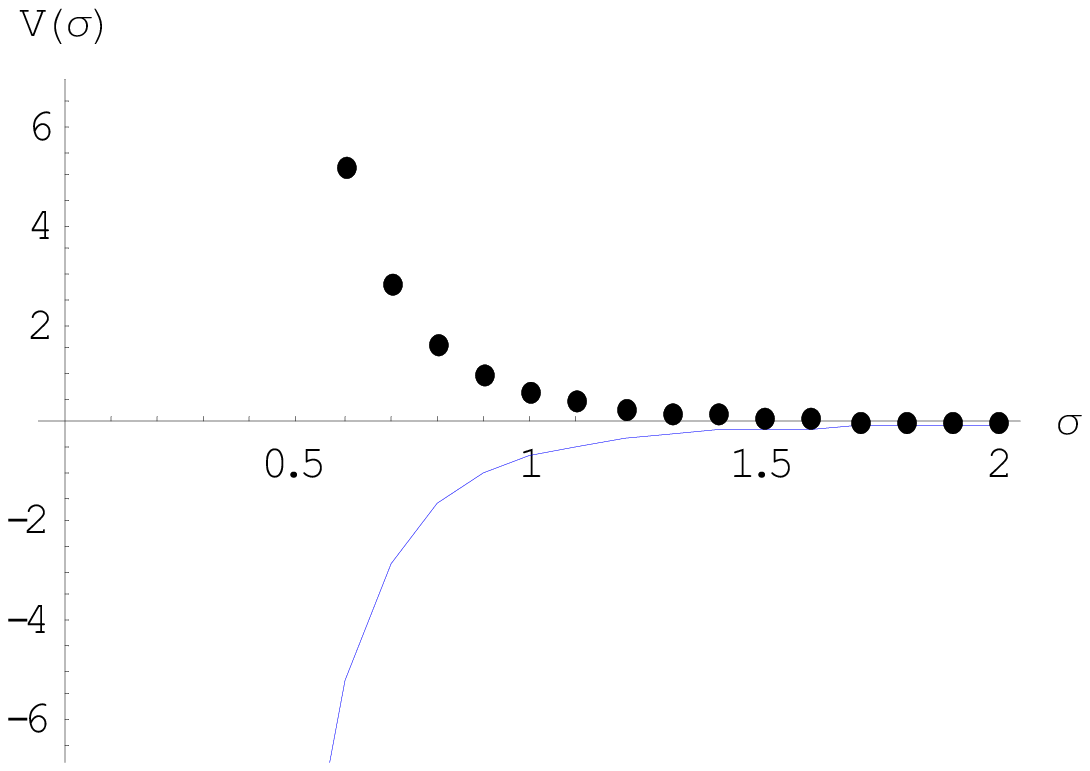}}
\end{center}
\bf{Figure 5 : }\rm The line shows the potential in zero mode of the relative funnel's fluctuations at the absence of electric field E and the dots represent the potential in the same mode at the presence of E. The presence of E is changing the potential totally to the opposite.
\vskip 1cm
\end{figure}

Now let's give the equation of motion of relative transverse
fluctuations of high $\ell$ modes with ($N-N_f$) strings
intersecting D3-branes. The fluctuation is given by $\delta \phi^i
(\sigma,t)=\sum\limits^{N-1}_{\ell=1}\psi^{i}_{i_1 ...
i_\ell}\al^{i_1} ... \al^{i_\ell} $ with $\psi^{i}_{i_1 ... i_\ell}$
are completely symmetric and traceless in the lower indices.

The action describing this system is \beq\bra{lll}
S&\approx-NT_1\int d^2\sigma  \Big[ (1-\lambda^2 E^2 )H-(1-\lambda
E)H\frac{\lambda^2}{2} (\partial_{t}\delta\phi^i)^2) \\\\
&+\frac{(1+\lambda E)\lambda^2}{2H} (\partial_\sigma \delta\phi^i)^2
-(1-\lambda E)\frac{\lambda^2}{2}\lbrack \phi^i ,\delta\phi^i
\rbrack^2 \\\\
&-\frac{\lambda^4}{12}\lbrack \partial_{\sigma}\phi^i
,\partial_{t}\delta\phi^i \rbrack^2+...\Big]. \era\eeq The equation
of motion for relative transverse fluctuations in high mode is as
follows

\beq \Big[\frac{1-\lambda E}{1+\lambda E}H\partial_{t}^2
-\partial_{\sigma}^2\Big]\delta\phi^i +(1-\lambda E)\lbrack \phi^i
,\lbrack \phi^i ,\delta\phi^i \rbrack\rbrack
-\frac{\lambda^2}{6}\lbrack \partial_{\sigma}\phi^i ,\lbrack
\partial_{\sigma}\phi^i ,\partial^2 _{t}\delta\phi^i
\rbrack\rbrack=0.\eeq

By the same way as done for overall transverse fluctuations the
equation of motion for each mode is \beq \Big[
-\partial_{\sigma}^2+\Big( \frac{1-\lambda E}{1+\lambda
E}(1+\lambda^2\frac{N^2 -1}{4\sigma^4})-\frac{\lambda^2
\ell(\ell+1)}{6\sigma^4}  \Big ) \partial_{t}^2 +(1-\lambda
E)\frac{\ell(\ell+1)}{\sigma^2}\Big]\delta\phi^i_\ell =0. \eeq We
take $\de\phi^i_\ell =f^i_\ell e^{-iwt}\de x^i$, then the equation
(41) becomes \beq \Big[ -\partial_{\sigma}^2 -\Big( \frac{1-\lambda
E}{1+\lambda E}(1+\lambda^2\frac{N^2 -1}{4\sigma^4})-\frac{\lambda^2
\ell(\ell+1)}{6\sigma^4}  \Big ) w^2 +(1-\lambda
E)\frac{\ell(\ell+1)}{\sigma^2}\Big]  f^i_\ell =0. \eeq To solve the
equation we choose for simplicity the boundaries of $\sigma$; For
small $\sigma$, the equation is reduced to \beq \Big[
-\partial_{\sigma}^2 -\Big( \frac{1-\lambda E}{1+\lambda
E}(1+\lambda^2\frac{N^2 -1}{4\sigma^4})-\frac{\lambda^2
\ell(\ell+1)}{6\sigma^4}  \Big ) w^2 \Big] f^i_\ell = 0, \eeq which
can be rewritten as follows \beq \Big[ -\frac{1+\lambda E}{1-\lambda
E}\partial_{\sigma}^2 -\Big( (1+\lambda^2\frac{N^2
-1}{4\sigma^4})-\frac{1+\lambda E}{1-\lambda E}\frac{\lambda^2
\ell(\ell+1)}{6\sigma^4}  \Big ) w^2 \Big] f^i_\ell = 0. \eeq We
change the coordinate to $\tilde{\sigma}=\sqrt{\frac{1-\lambda
E}{1+\lambda E}}w\sigma$ and the equation (44) becomes \beq \Big[
\partial_{\tilde{\sigma}}^2+ 1+ \frac{\kappa^2}{\tilde{\sigma}^4}
\Big] f^i_\ell (\tilde{\sigma}) = 0, \eeq with
$$\kappa^2=w^4\lambda^2 \frac{3(1-\lambda E)^2 (N^2
-1)-2(1-\lambda^2 E^2)\ell(\ell+1)}{12(1+\lambda E)^2}.$$ Then we
follow the suggestions of WKB by making a coordinate change; \beq
\beta(\tilde{\sigma})=\int\limits^{\tilde{\sigma}}_{\sqrt{\kappa}}dy\sqrt{1+
\frac{\kappa^2}{y^4}}, \eeq and \beq f^i_\ell
(\tilde{\sigma})=(1+\frac{\kappa^2}{\tilde{\sigma}^4})^{-\frac{1}{4}}\tilde{f}^i_\ell
(\tilde{\sigma}). \eeq Thus, the equation (45) becomes \beq\Big(
-\partial^{2}_{\beta}+V(\beta)\Big) \tilde{f^i}=0,\eeq with \beq
V(\beta)=\frac{5\kappa^2}{(\tilde{\sigma}^2
+\frac{\kappa^2}{\tilde{\sigma}^2})^3}. \eeq Then \beq
f^i_\ell=(1+\frac{\kappa^2}{\tilde{\sigma}^4})^{-\frac{1}{4}}e^{\pm
i\beta(\tilde{\sigma})}. \eeq The discussion is
similar to the overall case; so the obtained fluctuation has the
following limits; at large $\sigma$, $f^i_\ell\sim e^{\pm
i\beta(\tilde{\sigma})}$ and if $\sigma$ is small
$f^i_\ell=\frac{\sqrt{\kappa}}{\tilde{\sigma}}e^{\pm
i\beta(\tilde{\sigma})}$. These are the asymptotic
wave function in the regions $\beta\rightarrow \pm\infty$, while
around $\beta\sim 0$; i.e. $\tilde{\sigma}\sim\sqrt{\kappa}$,
$f^i_\ell\sim 2^{-\frac{1}{4}}$.\\

Then let's have a look at the potential in various limits of electric field;\\
\begin{itemize}
\item $E\sim \frac{1}{\lambda}$, $V(\beta)\sim 0$ \item $E\gg 1$,
$\kappa^2\equiv \kappa^2_+ \sim w^4\lambda^2 \frac{3(N^2
-1)+2\ell(\ell+1)}{12}$, then $\sigma\sim 0\Longrightarrow
V(\beta)\sim \frac{5\tilde{\sigma}^6}{\kappa^4_+}$ \item $E\ll 1$,
$\kappa^2 \equiv \kappa^2_- \sim w^4\lambda^2 \frac{3(N^2
-1)-2\ell(\ell+1)}{12}$; for this case we get $\sigma\sim
0\Longrightarrow V(\beta)\sim \frac{5\tilde{\sigma}^6}{\kappa^4_-}$
\end{itemize}
this means that we have a Neumann boundary condition with relative
fluctuations at small $\sigma$ (Fig.6).\\

Now, if $\sigma$ is too large the equation of motion (42) becomes
\beq \Big[ -\partial_{\sigma}^2 +(1-\lambda
E)\frac{\ell(\ell+1)}{\sigma^2}\Big] f^i_\ell =  \frac{1-\lambda
E}{1+\lambda E} w^2  f^i_\ell. \eeq

We see that the associated potential $V(\sigma)=(1-\lambda
E)\frac{\ell(\ell+1)}{\sigma^2}$ goes to $-\epsilon$ in the case of
$E\gg 1$ and to $+\epsilon$ if $E\ll 1$ since $\sigma$ is too large
with $\epsilon\sim 0$, (Fig.6). We get the same remark as before by dealing with the fluctuations for small and large $\sigma$ (50) and solving (51) respectively, at the presence of electric field that we have two separated regions depending on the electric field (fig.7).\\

We discussed quite explicitly through this section the fluctuation
of the funnel solution of D1$\bot$D3 branes by treating different
modes and different directions of the fluctuation. We found that the
system got an important property because of the presence of electric
field; the system has Neumann boundary condition.
\begin{figure}
\begin{center}
\mbox{\epsfysize=6cm
   \epsfxsize=6cm
  \epsffile{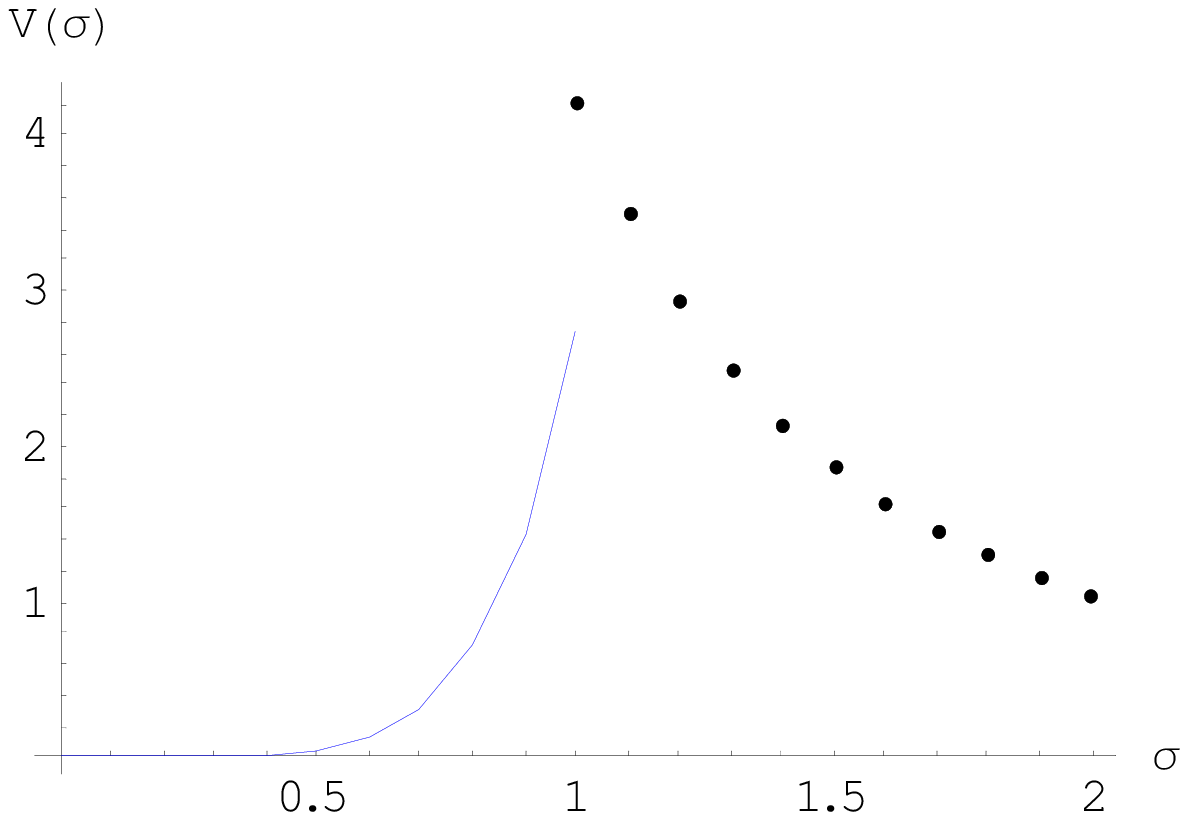}}
\mbox{\epsfysize=6cm
   \epsfxsize=6cm
  \epsffile{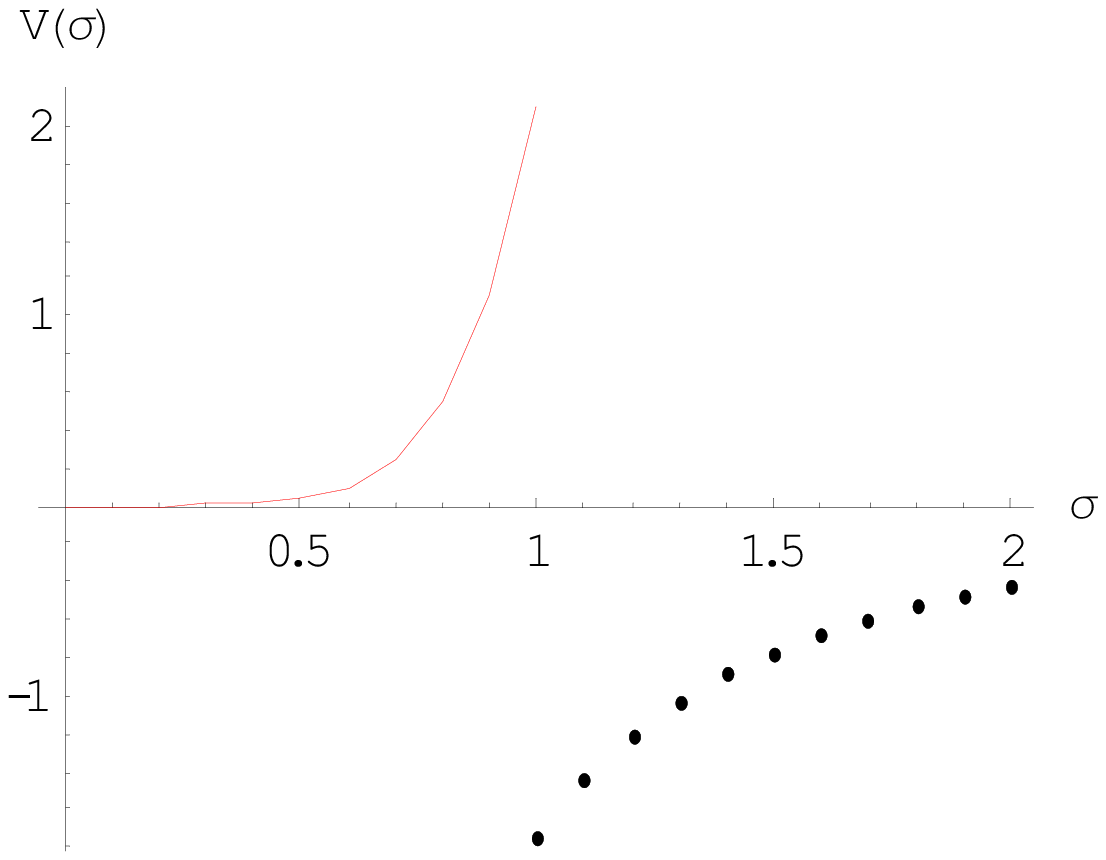}}
\end{center}
\bf{Figure 6 : }\rm As we saw in high mode of overall fluctuations, also for relative case we get high potential at some stage of $\sigma$ where the the tow curves meet representing potentials for small and large $\sigma$ at the absence of electric field E in the left figure. Right figure shows again a singularity this time in relative case because of the presence of E.
\vskip 1cm
\end{figure}
\begin{figure}
\begin{center}
\mbox{\epsfysize=6cm
   \epsfxsize=6cm
  \epsffile{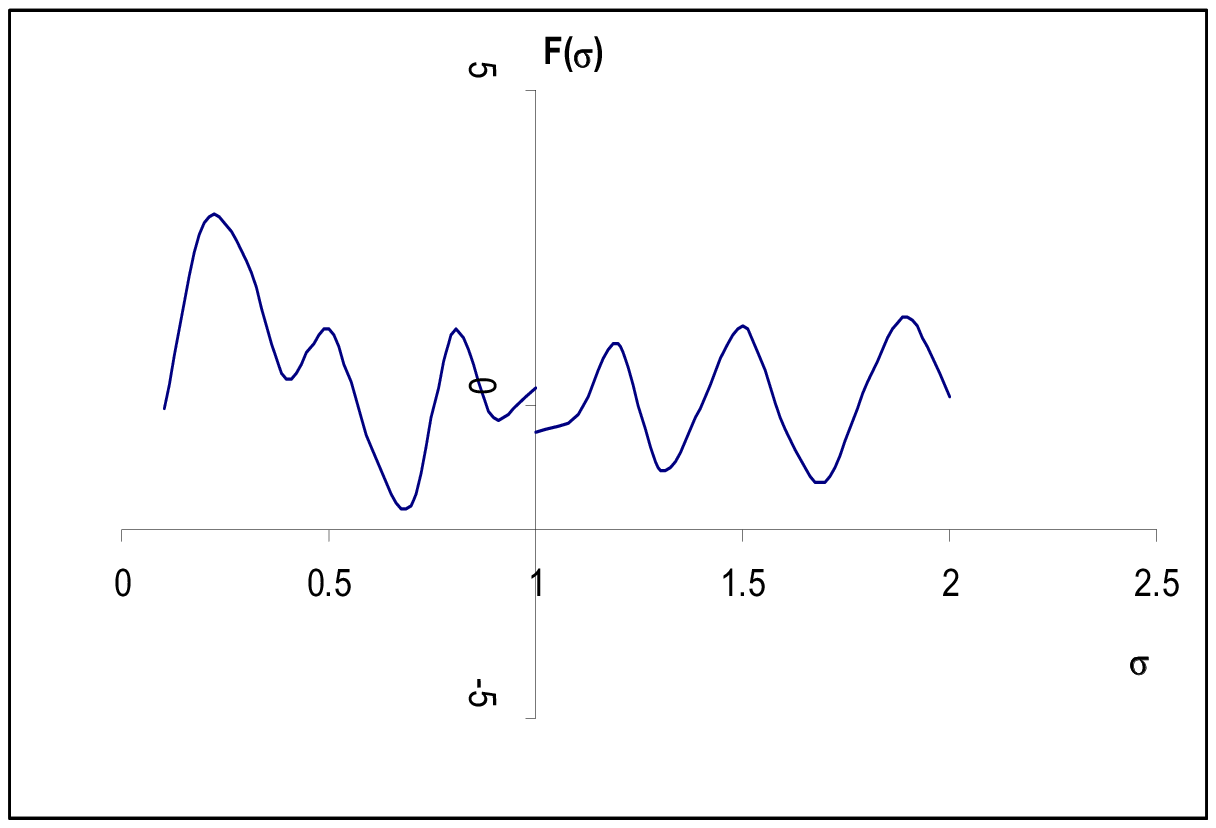}}
\end{center}
\bf{Figure 7 : }\rm The presence of electric field E causes a discontinuity of the wave in high mode of relative case meaning free boundary condition.
\vskip 1cm
\end{figure}
\section{Conclusion}
\hspace{.3in}We have investigated the intersecting D1-D3 branes
through a consideration of the presence of electric field. We have
treated the fluctuations of the funnel solutions and we have
discussed explicitly the potentials in both systems. We found a
specific feature of the presence of electric field. When the
electric field is going up and down the potential of the system is
changing and the fluctuations of funnel solutions as well which
cause the division of the system to tow regions. Consequently, the
end point of the
dyonic strings move on the brane which means we have Neumann boundary condition.\\

The present study is in flat background and there is another
interesting investigation is concerning the perturbations
propagating on a dyonic string in the supergravity background
\cite{supergravComp,dual} of an orthogonal 3-brane. Then we can deal
with this important case and see if we will get the same boundary
conditions by treating the dyonic fluctuations.

\section*{Acknowledgements}
\hspace{.3in}This work was supported by a grant from the Arab Fund
for Economic and Social Development.


\begin{thebibliography}{99}
\bibitem{BI} J. Polchinski, Tasi Lectures on D-branes,
hep-th/9611050; R. Leigh, Mod. Phys. Lett .A4 (1989) 2767.

\bibitem{InterBran1} C.G. Callan and J.M. Maldacena, Nucl. Phys.
B513 (1998) 198, hep-th/9708147.

\bibitem{InterBran2} G.W. Gibbons, Nucl. Phys. B514
(1998) 603, hep-th/9709027; P.S. Howe, N.D. Lambert and P.C. West,
Nucl. Phys. B515(1998) 203, hep-th/9709014; T. Banks, W. Fischler,
S. H. Shenker and L. Susskind, Phys. Rev. D55 (1997) 5112,
hepth/9610043; D. Kabat and W. Taylor, Adv. Theor. Math. Phys. 2
(1998) 181, hep-th/9711078; S. Rey, hep-th/9711081; R.C. Myers, JHEP
9912 (1999) 022, hep-th/9910053.

\bibitem{fun} D. Brecher, Phys. Lett. B 442 (1998) 117,
hep-th/9804180; P. Cook, R. de Mello Koch and J. Murugan, Phys. Rev.
D 68 (2003) 126007, hep-th/0306250; J. K. Barrett and P. Bowcock,
hep-th/0402163.

\bibitem{dual}N. R. Constable, R. C. Myers and O. Tafjord, Phys.
Rev. D61, 106009 (2000), hep-th/9911136.

\bibitem{fluct} S.-J. Rey and J.-T. Yee, Nucl. Phys. B52 (1998)
229, hep-th/9711202; S. Lee, A. Peet and L. Thorlacius, Nucl. Phys.
B514 (1998) 161, hep-th/9710097; D. Kastor and J. Traschen, Phys.
Rev. D61 (2000) 024034, hep-th/9906237; S.-J. Rey and J.-T. Yee,
Eur. Phys. J. C22 (2001) 379, hep-th/9803001.

\bibitem{fuzfun} R. Bhattacharyya and R. de Mello Koch,
hep-th/0508131; C. Papageorgakis, S. Ramgoolam, N. Toumbas, JHEP
0601 (2006) 030, hep-th/0510144; D. Bak, J. Lee and H. Min, Phys.
Rev. D59 (1999) 045011, hep-th/9806149.

\bibitem{funnelSoluD5}J. Castelino, S. Lee and W. Taylor, Nucl.
Phys. B526, 334 (1998), hep-th/9712105; H. Grosse, C. Klimcik and P.
Presnajder, Commun. Math. Phys. 180, 429 (1996), hepth/9602115.

\bibitem{NBCBI}K. G. Savvidy and G. K. Savvidy, Nucl. Phys. B561 (1999) 117,
hep-th/9902023.

\bibitem{cm} N. R. Constable, R. C. Myers, O.
Tafjord, JHEP 0106 (2001) 023, hep-th/0102080.

\bibitem{Gib} G. W. Gibbons, Nucl.Phys. B514 (1998) 603-639, hep-th/9709027.

\bibitem{supergravComp}S. Lee, A. Peet and L. Thorlacius, Nucl.
Phys. B514, 161 (1998), hep-th/9710097; D. Kastor and J. Traschen,
Phys. Rev. D61, 024034 (2000), hep-th/9906237.


\end{thebibliography}
\end{document}